\documentclass[10pt,aps,prl,twocolumn,preprint]{revtex4}   

\usepackage{amsmath}    
\usepackage{graphicx}   

\begin{document}

\title{Electromagnetic wave scattering by a superconductor}
\author{Miguel C. N. Fiolhais}
\email{miguel.fiolhais@cern.ch}   
\affiliation{LIP, Department of Physics, University of Coimbra, 3004-516 Coimbra, Portugal, EU}
\author{Hanno Ess\'en}
\email{hanno@mech.kth.se}   
\affiliation{Department of Mechanics, KTH, 10044 Stockholm, Sweden, EU}

\date{\today}

\begin{abstract}
The interaction between radiation and superconductors is explored in this paper. In particular, the calculation of a plane standing wave scattered by an infinite cylindrical superconductor is performed by solving the Helmholtz equation in cylindrical coordinates. Numerical results computed up to $\mathcal{O}(77)$ of Bessel functions are presented for different wavelengths showing the appearance of a diffraction pattern.

The following article is published in Europhysics Letters: http://iopscience.iop.org/0295-5075/97/4/44006/.

\end{abstract}

\maketitle

\section{Introduction}
Since the discovery of magnetic field expulsion from superconductors, commonly known as Meissner effect, by Meissner and Ochsenfeld in 1933 \cite{meissner}, the solutions of Maxwell equations for a superconductor in the presence of external fields have been analyzed in numerous occasions for different geometries \cite{reitz,matute,zhilichev,batygin,fiolhais}. While most of these studies are focused on static external magnetic fields, the interaction between superconductors and time-dependent electromagnetic fields remains unexplored in depth in the scientific literature. 

In this paper we try to change this situation by solving the problem of a standing wave scattered by an infinite cylindrical superconductor. The problem is addressed by starting from the time-dependent Maxwell equation for the source-free case (wave equation) outside the superconducting region. For a standing wave, the space and time components of the wave equation solution are independent, reducing it to the Helmholtz and simple harmonic motion equations. Finally, a linear combination of cylindrical Bessel functions (general solutions of the Helmholtz equation) is presented as the final solution of the problem respecting the boundary conditions at infinity and at the superconductor surface.
The main difficulties in problems like this may arise from the symmetries of the system and boundary condition problem. Unlike the cylindrical case where the cylindrical symmetry simplifies the problem, in other geometries such as the spherical, the problem becomes non-trivial.

To the best of our knowledge, the only extensive study related to the subject in the linear domain was performed by Nye in 2003 \cite{nye} where a detailed analysis of the scattering of plane electromagnetic waves by perfectly conducting regions is presented. The results are consistent with ours. Other works addressing the interaction between
electromagnetic waves and nonlinear superconducting materials have been presented mainly within the microwave community. A complete study with numerical computations can be found in \cite{caorsi}.

\section{Standing wave scattering by an infinite cylindrical superconductor}
\begin{figure}[h]
\begin{center}
\includegraphics[height=4.5cm]{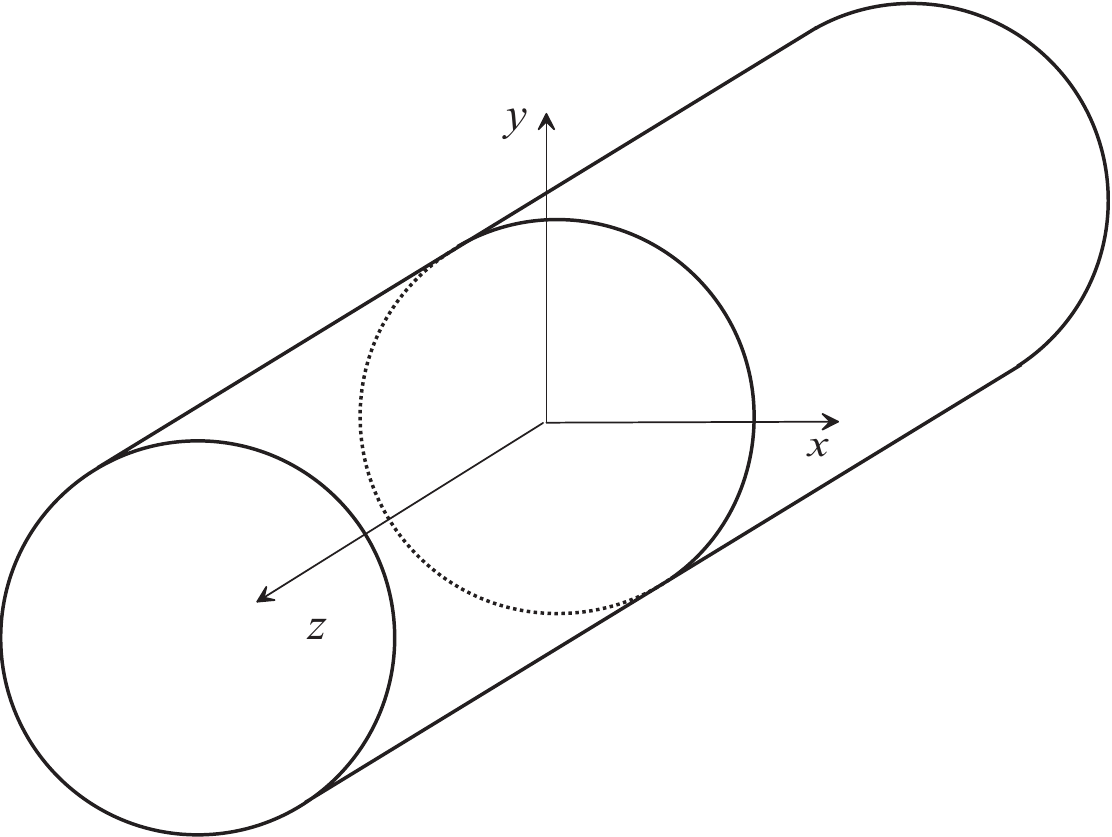}
\caption{Infinite cylinder section.}
\label{cylinder}
\end{center}
\end{figure}

Consider an infinite cylindrical superconductor with radius $R$ in the presence of a plane standing electromagnetic wave and that the cylinder axis coincides with the z-axis (see Figure \ref{cylinder}).

The Maxwell equation:
\begin{equation}
\nabla \times \mathbf{B} = \mu_0\mathbf{J} + \frac{1}{c^2} \frac{\partial \mathbf{E}} {\partial t},
\end{equation}
in the source-free case and assuming the relations $\mathbf{B} = \nabla \times \mathbf{A}$ and $\mathbf{E} = - \nabla \phi -  \frac{\partial \mathbf{A}} {\partial t}$, becomes:
\begin{equation}
\nabla \times \left ( \nabla \times \mathbf{A} \right ) + \frac{1}{c^2} \frac{\partial^2 \mathbf{A}} {\partial t^2} = 0,
\end{equation}
which can be simplified by taking the Lorentz gauge ($\nabla \cdot  \mathbf{A} + \frac{1}{c^2} \frac{\partial \phi} {\partial t} = 0$):
\begin{equation}
 \nabla^2 \mathbf{A} - \frac{1}{c^2} \frac{\partial^2 \mathbf{A}} {\partial t^2} = 0,
\end{equation}
the wave equation. If the electromagnetic wave propagates from infinity in the x-axis direction with the electric field pointing in the direction of the cylinder axis, the z-axis component of the vector potential $\mathbf{A}$ is the only relevant component needed to compute the electromagnetic field:
\begin{equation}
 \nabla^2 A_z - \frac{1}{c^2} \frac{\partial^2 A_z} {\partial t^2} = 0.
 \label{wave}
\end{equation}
As the superconducting cylinder is in the presence of a plane standing electromagnetic wave, the component $A_z$ in cylindrical coordinates (since the cylinder is infinite there are no dependence on the z-coordinate) can be separated into: $A_z = F(r,\phi)T(t)$, a space-like function $F(r,\phi)$ and a time-like function $T(t)$. This separation of variables gives rise to two equations:
\begin{equation}
\nabla^2 F(r,\phi)+ \frac{1}{c^2} F(r,\phi) = 0
\label{helmholtz}
\end{equation}
and
\begin{equation}
 \frac{\textrm{d}^2 T(t)} {\textrm{d} t^2} + {\omega^2} T(t) = 0,
\label{SHM}
\end{equation}
the Helmholtz equation \ref{helmholtz} and the simple harmonic motion equation \ref{SHM}. Therefore, the most general solution for $r>R$ is:
\begin{eqnarray}
A_z (r,\phi,t) = &\sum_{m=0}^\infty &  [A_m\cos(m\phi)+B_m\sin(m\phi)] \nonumber \\
                          & \times                      & [C_mJ_m(kr)+D_mY_m(kr)] \nonumber \\
                          & \times                      & [E\cos(\omega t )+F\sin(\omega t )].
\label{general}
\end{eqnarray}
$J_m(kr)$ and $Y_m(kr)$ are the cylindrical Bessel functions and $k = \frac{2\pi}{\lambda}$, where $\lambda$ is the wavelength.
The next step is to find the particular solution that satisfies the boundary conditions of a plane standing wave coming from infinity towards the superconducting cylinder.

The superconducting cylinder is considered to be of type I with zero penetration length, a reasonable approximation if the cylinder radius is much larger than the penetration length. The magnetic field is expelled from the superconducting region by the surface current $\mathbf{k}_S$ defined in the coupling equation between the superconductor and the magnetic field:
\begin{equation}
\mathbf{k}_S  = \frac{c}{4\pi} \hat{n} \times \mathbf{B^+},
\label{coupling}
\end{equation}
where $\hat n$ is the unitary vector orthogonal to the superconductor surface and $\mathbf{B^+}$ is the magnetic field in the outer region of the superconductor surface. As the magnetic field and currents are zero inside the superconducting cylinder, the component $A_z$ must be constant inside (set to zero for convenience).  As there cannot be any angular dependencies on the surface of the cylinder, then $D_m=-C_m\frac{J_m(kR)}{Y_m(kR)}$ and:
\begin{eqnarray}
A_z (r,\phi,t) = &\sum_{m=0}^\infty &  [A_m\cos(m\phi)+B_m\sin(m\phi)] \nonumber \\
                          & \times                      & [C_mJ_m(kr)-C_m\frac{J_m(kR)}{Y_m(kR)}Y_m(kr)] \nonumber \\
                          & \times                      & [E\cos(\omega t)+F\sin(\omega t)].
\label{generalR}
\end{eqnarray}

At infinity ($r\rightarrow \infty$), the solution must take the form of the plane standing wave assumed to be:
\begin{eqnarray}
A_z (r,\phi,t) = A_0 \cos(k r\cos\phi) \cos(\omega t),
\label{plane}
\end{eqnarray}
where $A_0$ is the amplitude of the wave and $x= r \cos\phi$. According to Jacobi-Anger expansion \cite{abramowitz},

\begin{equation}
\cos(z \cos \phi) = J_0(z)+2 \sum_{m=1}^{\infty}(-1)^m J_{2m}(z) \cos(2m \phi),
 \end{equation}
and therefore the final solution reads for $r>R$:
\begin{eqnarray}
F (r,\phi)                     & =                             & J_{0}(kr)-\frac{J_{0}(kR)}{Y_{0}(kR)}Y_{0}(kr) \nonumber \\
                                           & +                             &2\sum_{m=1}^\infty   (-1)^{m} \cos(2m\phi) \nonumber \\
                                           & \times                     & \left[J_{2m}(kr)-\frac{J_{2m}(kR)}{Y_{2m}(kR)}Y_{2m}(kr) \right], 
\label{final}
\end{eqnarray}
and $F (r,\phi) = 0$ for $r\leq R$. Such a mathematical result is hard to interpret and visualize but possible to represent with enough accuracy through numerical computations up to $m=77$. The spatial representation and analysis of $A_z$ for different wavelengths are presented in the following section.

\section{Simulation}

\begin{figure}[ht]
\begin{center}
\includegraphics[height=8.cm]{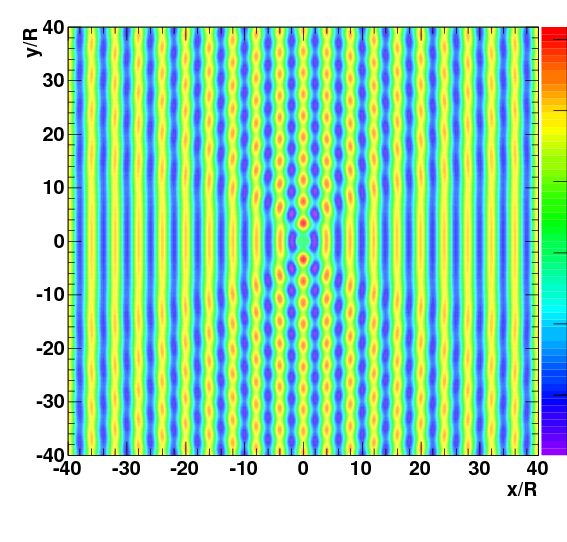} \\
\includegraphics[height=8.cm]{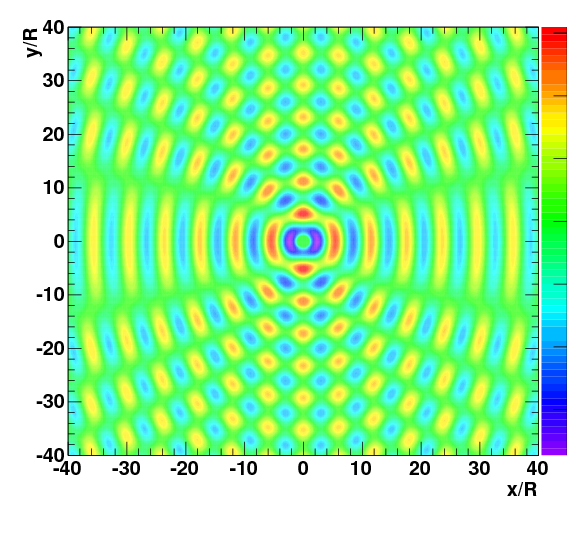}
\caption{Standing wave scattered around the infinite cylindrical superconductor for $\lambda = 4 R$ (top) and $\lambda = 6 R$ (bottom).}
\label{planewave}
\end{center}
\end{figure}

The numerical computations that simulate the scattering were performed in a 2D array representing the cross-section of the superconducting cylinder (centered at the origin) in the x-y plane ranging from minus 40 to 40 in steps of 0.1 radius units for both $x$ and $y$ directions at the instant $t=0$. The simulation of the plane standing wave scattering by the superconducting cylinder is shown in Figure \ref{planewave} in arbitrary units for $\lambda = 4 R$ (top) and $\lambda = 6 R$ (bottom). For $\lambda = 4 R$ the scattering presents a diffraction pattern of nodes and anti-nodes around the superconducting cylinder while converging to the plane wave at large distances. The scattering effects are perhaps more visible for $\lambda = 6 R$, the nodes and anti-nodes around the cylinder are clearly amplified and separated by roughly the wavelength. The nodes and anti-nodes are maximal near the superconductor and decrease gradually with distance.

\section{Discussion}

According to Maxwell equations, the interaction between a plane wave and a superconductor gives rise to a diffraction pattern. We would like to challenge experimental groups to test the result obtained in this paper even though we are aware of some technical limitations that may constrain the observations. For example, a difficulty may come from generating a large enough controlled plane standing wave with a wavelength of the same order of magnitude as the superconductor radius to enhance the scattering. Also, it is important to stress that the wave period must be much larger than the superconductor relaxation time so that it can respond instantly to field variations.

\end{document}